# Alternative Agriculture Land-Use Transformation Pathways by Partial-Equilibrium Agricultural Sector Model: A Mathematical Approach

**Malvika Kanojia[1], Prerna Kamani[2], Gautam Siddharth Kashyap[3], Shafaq Naz[4], Samar Wazir[5], Abhishek Chauhan[6]**

[1]malvika000017@gmail.com, [2]prernakemani@gmail.com, [3]officialgautamgsk.gsk@gmail.com, [4]shafaq.2017.naz@gmail.com, [5]samar.wazir786@gmail.com, [6]abhi7.chauhan96@gmail.com

[1]F-131/132, Near Ram Mandir, West Vinod Nagar, New Delhi, India
[2]B-74-S/4 Ganapati Apartment, Dilshad Colony, Delhi, India
[3,4,5]Department of Computer Science and Engineering, SEST, Jamia Hamdard, New Delhi, India
[6]31 Country Homes West South City, Ludhiana, Punjab, India

**Abstract**

Humanity's progress in combating hunger, poverty, and child mortality is marred by escalating environmental degradation due to rising greenhouse gas emissions and climate change impacts. Despite positive developments, ecosystems are suffering globally. Regional strategies for mitigating and adapting to climate change must be viewed from a global perspective. The 2015 UN Sustainable Development Goals reflect the challenge of balancing social and environmental aspects for sustainable development. Agriculture, vital for food production, also threatens Earth systems. A study examines the interplay of land-use impacts, modeling crop and livestock trade, and their effects on climate, biodiversity, water, and land using a Partial-Equilibrium Agricultural Sector Model. Different scenarios involving taxing externalities related to Earth processes were tested. Results show synergies in reducing emissions, biodiversity loss, water use, and phosphorus pollution, driven by shifts in crop management. Nitrogen application and deforestation scenarios exhibit weaker synergies and more conflicts. The study offers insights into SDG interactions and the potential for sustainable farming.

## 1. Introduction

Despite persistent inequities, there is a global trend toward better living for all, as seen by the Human Development Index (HDI) steady rise and the subsequent declines in multidimensional poverty and infant mortality [1]. Overall, considerable progress has been made in human development, particularly over the past century. It has also had a significant impact on the environment at the same time. An unprecedented rate of climate change is being caused by the release of anthropogenic GHG. The sixth mass extinction, which is the first to be directly ascribed to humans, is being caused by widespread land-use change and invasions into the few remaining unaltered habitats. These and other anthropogenic effects have accumulated to the point where some believe a new geological epoch, the Anthropocene [2], named after humans as the dominant force responsible for reshaping the world, has begun. According to Rockstrom et al. [3], human influence may also result in "abrupt environmental change within continental- to planetary-scale systems" in addition to this ongoing environmental degradation. More specifically, [3] have estimated the "safe operating space for humans for the functioning of the Earth system" by identifying nine biophysical thresholds, also known as planetary boundaries, for important Earth system processes. They propose that to maintain the relative environmental stability that mankind has so far experienced and benefited from, human impact on the related subsystems and processes should be kept to a minimum. The Earth system may undergo irreversible and quick changes if these boundaries are not upheld, which will probably be detrimental to most or all of humankind and all species on Earth. Even if adaptation to such changes is conceivable, doing so would probably be expensive and detrimental to human growth. As a result of their increased exposure and vulnerability, less developed areas may be particularly seriously affected [4]. Because of this, the 2015 UN-SDGs for 2030 place a strong emphasis on both humanity's upward trajectory and a sustainable way of living within the bounds of the planet. The UN-2017 does not specifically mention the idea of planetary boundaries, but achieving the SDGs is inextricably tied to them. This is particularly true for the three SDGs that directly address the environment, SDGs 13, 14, and 15: life on land, life below the sea, and climate action. SDG-13 naturally relates to a necessary global cap on GHG emissions and anthropogenic climate change, as stated by [3]. The limits on ocean acidity, which are rising as a result of $CO_2$ emissions, and flows of nitrogen and phosphorus into water bodies, which have an impact on marine ecosystems, are directly related to SDG-14. SDG-15 is directly related to land use change, particularly through deforestation and the expansion of farmland, chemical soil contamination, and biodiversity loss. Additionally, the limit on freshwater consumption globally is connected to SDG-6 on clean water and sanitation.

Additionally, SDG-2—zero hunger—is linked to all of these restrictions. This is because large-scale agriculture is essential to ensuring healthy lives and essential to fulfilling SDG-2's goal of food security. Land use has a detrimental impact on all Earth system processes connected to planetary boundaries at the same time [5].

In contrast, there is not a lot of prior research on the connections, overlaps, and conflicts between various sets of SDGs and planetary boundaries. The Earth3 model by Randers et al. [6], which aims to model sustainable development trajectories in line with all SDGs and within all planetary limitations, is an exception. Since it is imperative to respect all planetary boundaries, further study in this area is required. A greater understanding of the links between the associated Earth system processes is crucial to shaping evidence-based routes into a more sustainable future. Our understanding of the connections between several Earth system processes related to planetary boundaries and impacted by the agricultural sectors—an important sub-sector of human development—will be advanced by this work. Agriculture is in the challenging position of balancing multiple human needs and SDGs. In this study, the author specifically develop a PEASM to calculate several sectoral sustainability indicators related to planetary boundaries. The author attempt to respond to the question i.e. "How strong are interactions and conflicts among various effects of land use on the world, in particular between climate change, biosphere integrity, nitrogen and phosphorus biogeochemical cycles, freshwater use, and land system change?" by executing several scenarios, forcing each indicator to its attainable minimum within predetermined limitations, and analysing the shifting levels of all the indicators. The author creates the model using the General Algebraic Modeling System (GAMS), which is frequently used for agricultural optimization problems and models, such as the Model of Agricultural Production and its Impact on the Environment (MAgPIE), the Global Biosphere Management Model (GBMM), and the European Forest and Agricultural Sector Optimization Model (EU-FASOM). The model is resolved by GAMS as a linear programming issue using the commercial CPLEX solver. The agricultural impacts cannot be fairly estimated by the model to be near or far from each border. However, within the predetermined limits, the author can use it to gauge correlations and gauge the relative change in these indicators. These findings and this model can then serve as the foundation and framework for further, in-depth research.

This paper is organised as follows; in the part after that, we'll look at how it relates to the strategy we've suggested. The comprehensive mathematical model is described in section 3. The results of the suggested mathematical model are explained in section 4. he paper's opinions and a discussion of the research issue are offered in section 5, and section 6 serves as the paper's conclusion.

## 2. Literature Review

In light of the suggested study, some works that are aligned with ours include a PEASM research of Mediterranean climate change that has been proposed by Palatnik et al. [7]. The main objective is to quantify and evaluate the socioeconomic effects of anticipated medium-term climate change. As a result, the economic framework more accurately depicts the variety of facts regarding land output. The trade-offs or synergies between mitigation and food security in agricultural output and GHG emissions have been researched by Valin et al. [8]. They've looked into how land use change and GHG emissions in developing countries are affected by various crop production and animal feed efficiency scenarios. They illustrated the importance of the rebound effect using a model of economic equilibrium. Mosnier [9] used a global partial equilibrium model to investigate monitoring the indirect effects of climate change adaptation and mitigation strategies for agriculture and land use change. They asserted that the Global Partial Equilibrium Agricultural and Forestry Model (GLOBIOM) is used to assess the risks associated with climate change adaptation and mitigation strategies that leak through global trade. The impact of agricultural-based biofuel production on GHG emissions brought on by land-use change has been hypothesised by Panichelli and Gnansounou [10]. On the other hand, model integration calls for unified temporal and scale techniques to provide generic representations of the system. They have concluded that connecting the results of many models is a useful strategy for handling effects on various time and space scales. The global consequences of first- and second-generation biofuel objectives for land usage have been put forth by Havlik et al. [11]. According to their calculations, second-generation biofuel production fueled by wood from sustainably managed existing forests will provide negative $i$(land-use change) factor, resulting in 27% lower global emissions by 2030 than the "No biofuel" scenario. They advocate for legislative action that focuses less on the actual production of biofuels and more on the positive and negative effects that it has on the environment and society. We will see the suggested mathematical model in the section after this.

## 3. Mathematical Models

By maximising agricultural welfare and taxing several harmful externalities, the author evaluates the positive and negative effects of reducing agricultural pressure on the processes of the studied Earth system as:

$$\text{Max} \quad W = \sum_{t,r,p}(z_{r,p}.Q_{t,r,p}) - \sum_{t,r,c,m}(c_{r,c}^c.c_m^m.C_{t,r,c,m}) - \sum_{t,r,l}(c_{r,l}^l.L_{t,r,l}) - \sum_{t,r,q}(c_{r,q}^q.P_{t,r,q}) - \sum_{t,r',r,p}(z_{r,p}.0.15.T_{t,r',r,p}) -$$
$$\sum_{t,r,p}(z_{r,p}.(0.01.(S_{t,r,p}^+ + S_{t,r,p}^-) + 0.03.S_{t,r,p}^L)) - \sum_{t,r,e}(t_e.E_{t,r,e}) \quad (1)$$

*where* uppercase letters represent non-negative variables defined by the model (except for *W*), lowercase letters represent parameters supplied into the model, and subscripts represent indices to be summed across. Going from left to right, *W* represents the total welfare (1000 USD) created. $Q_{t,r,p}$ is the quantity (1000 t) of any consumable good *p* sold in region *r* and period *t* at price *z* (USD t$^{-1}$), which remains consistent over time for every area *r* and product *p*; $C_{t,r,c,m}$ is the area (1000 ha) used in any region *r* and period *t* to grow crop *c*, utilizing management system *m*. $c_{r,c}^c$ is the time-independent cost (USD ha$^{-1}$) per crop *c* in region *r*, which is modified through the cost factor $c_m^m$, which is determined by the management system *m*; $L_{t,r,l}$ is the amount of animals *l* (1000 heads) raised in region *r* and period *t*, $c_{r,l}^l$ the associated time-independent cost; $P_{t,r,q}$ is the amount (1000 t) of input products processed in process *q*, region *r* and period *t* at cost ($c_{r,q}^q$ (USD/input t); $T_{t,r',r,p}$ is the amount (1000 t) of products *p* shipped internationally from exporter region *r'* to the consuming or re-trading region *r* in period *t*; $S_{t,r,p}^L$ is the storage level (1000 t), $S_{t,r,p}^+$ and $S_{t,r,p}^-$ is the amount of product *p* added to and taken from storage in region *r* and period *t*. Finally, $E_{t,r,e}$ summarizes the extent of the externalities *e* in region *r* and period *t* taxed at a rate $t_e$ (USD/unit externality) in the model. The underlying assumptions, restricting equations, and data inputs will be discussed further in the subsequent sections of this section.

The model imitates a fundamental global production, consumption, and commerce network using this objective function. This method uses a range of minimum and maximum calorie intake per person to control consumption. The amount of land that can be used for crop growth and animal grazing, the maximum yearly land use change, and historical land use, crop, and livestock-mix data all have an impact on production. The model includes 33 regions, including the European Unions (EU's) 28 member states on a national level as well as 5 combined RoW regions, including Latin America (LAM), Northern America (NAM), Central, South, and East Asia (ASI), Africa and West Asia (MAF), and significant portions of Oceania (OCE). As a result, it leaves out non-EU Europe, such as Russia and many small island nations. The author made this decision to promote efficient data extraction and alignment, primarily from and between Food and Agriculture Organization Corporate Statistical Database (FAOSTAT) and the Shared Socioeconomic Pathways (SSP) database i.e. SSPDB, of the International Institute for Applied System Analysis.

To calculate parameters for these RoW zones, the author often adds up or average (without weight) data for subregions. The author uses data for a subset of regional nations in the few instances where there are no sub-regional data available. Between 2015 and 2050, there are 5-year intervals in the temporal resolution. The 34 products that are covered by crop agriculture, livestock production, or product processing are listed in Table 1. According to Food and Agriculture Organization (FAO) figures for 2015, the author selected this option due to its global or regional significance, as these products either make up a significant portion of the calories consumed or are consumed in huge amounts. Based on land use, management practices, and land-use change, we calculate several sustainability impact indicators for the examined Earth system processes, including GHG emissions from crop and livestock production, land-atmosphere carbon cycling, land-use change, and international trade; blue and grey freshwater use from crop production; land-use change from deforestation, afforestation, and total cropland area; and BDL from land use. The author studies several scenarios in which a separate negative externality is internalised via taxes. The author compares the resulting indicators for each of these scenarios to a basic scenario with no taxes. GHG emissions (t $CO_{2\,eq}$), BDL (1000 ha), synthetic nitrogen fertiliser application (t), phosphorus application (t), blue and grey freshwater usage (1000 m$^3$), and deforestation are all chargeable externalities (1000 ha).

Table 1: Products that the model covers

| | |
|---|---|
| **Livestock Primary** | Eggs, pork, beef, milk, poultry |
| **Crop Primary** | bananas, apples, barley, rapeseed, rice, sorghum, soybeans, beans, cassava, chickpeas (and other pulses), maize, millet, oil palm fruits*, olives, onions, oranges, potatoes, sugarcane*, sunflower seeds, tomatoes, wheat |
| **Process Products** | rapeseed oil, olive oil, palm oil, skimmed milk, butter, feed*, soy oil, sunflower oil |

* Denotes products not meant for immediate human consumption.

### 3.1 Consumption

Consumption $Q_{t,r,p}$, being the sole positive component in Equation (1), is constrained by the availability of consumable items as well as the caloric need in any place:

$$n_{t,r} \cdot 1.1 \leq \frac{\sum_p (Q_{t,r,p} \cdot k_p)}{p_{t,r}} \leq n_{t,r} \cdot 1.8 \quad \forall t, r, \tag{2}$$

where $p_{t,r} = \sum_{a,g}(p_{t,r,a,g})$ (3)

And *where* $n_{t,r}$ is the Minimum Dietary Energy Requirement (MDER), of the population in region $r$ in period $t$ and $k_p$ is the average amount of energy (1000 kcal $t^{-1}$) contained in consumable product $p$. The author chooses factors 1.1 and 1.8 to reflect unequal food distribution and food loss on the lower and wealthy consumption behaviour on the upper end. $p_{t,r}$ is the total population, $p_{t,r,a,g}$ the number of people (millions) in age group $a$ and of sex $g$ in region $r$ and period $t$, as projected in the SSPDB for SSP 2 [12]. The author estimate $n_{t,r}$ for each region $r$ and period $t$ as a weighted average, as:

$$n_{t,r} = \frac{\sum_{a,g}(n_{a,g} \cdot p_{t,r,a,g})}{\sum_{a,g} p_{t,r,a,g}} \cdot 365 \tag{4}$$

where $n_{a,g}$ is the estimated daily calorie need per person of age $a$ and sex $g$, based on recommendations for moderately active American citizens. Even while Equation (3) guarantees that the MDER for every person can be fulfilled even without corrective factors, producing only exactly $n_t^{global}$ a month would probably suggest widespread undernourishment because of things like food loss and unequal distribution [13]. The model estimates that a 10% increase in consumption would be roughly necessary to achieve SDG-2 while considering these problems. Significantly greater production may be required to make up for current distribution disparities. On the other end of the spectrum, in wealthy nations where hunger is not a significant issue, food consumption is not limitless. According to FAO statistics, the supply in the most affluent EU countries, like Austria, Belgium, or Ireland, can reach as high as 3700–3900 kcal per person per day. With a global average MDER of 2100 and the assumption that they represent the highest limit of typical consumption, the top limit is calculated to be 1.8. As shown in Equation (5), the author estimates the caloric content per tonne of product $p$, $k_p$.

$$k_p = \sum_{r'} \left( \frac{g_{r',p} \cdot 365}{m_{r',p}} \right) \cdot \frac{1000}{\sum_{r'} 1} \tag{5}$$

where $g_{r',p}$ is the food supply (kcal per capita$^{-1}$ day$^{-1}$) and $m_{r,p}$ the food supply quantity (kg capita$^{-1}$ year$^{-1}$) of product $p$ in region $r'$. While the author calculates $k_p$ for global use, $r'$ is a subset of $r$ depending on the specific product $p$. This varies, as some regions may show $m_{r,p} = 0$ for some products in the dataset. Furthermore, while Equation (5) is straightforward in principle, for some regions and products, the data produces $k_p$ many times higher than reference values from other sources, such as the U.S. Department of Agriculture [14]. To filter these outliers, the author only considered region results for a product if $0 < \frac{g_{r',p} \cdot 365}{m_{r',p}} < 9000$ kcal kg$^{-1}$, resulting in global estimates as shown in Table 2.

Table 2: Estimated number of calories per product tonne (kcal kg-1)

| Food | kcal | Food | kcal | Food | kcal |
|---|---|---|---|---|---|
| Barley | 2591 | Sunflower Oil | 8811 | Beans | 3400 |
| Butter | 7346 | Soy Oil | 8835.4 | Apples | 539 |
| Eggs | 1419.4 | Skimmed Milk | 209.2 | Feed | - |
| Milk | 547 | Rapeseed | 2255 | Cassava | 1488.2 |
| Olives | 1100 | Pork | 2100 | Olive Oil | 8805.3 |
| Oranges | 403.2 | Bananas | 601.4 | Palm Oil | 8809.6 |
| Potatoes | 675.3 | Beef | 1445.1 | Rice | 2022.4 |
| Rapeseed Oil | 8600.1 | Chickpeas | 3567.6 | Poultry | 1337 |
| Sorghum | 3166 | Maize | 2922 | Wheat | 2835.6 |
| Sugarcane | - | Oil palm fruit | - | Sunflower seeds | 3115.2 |
| Tomatoes | 233.1 | Onions | 380 | Soybeans | 2744.3 |
| Wheat | 8819.6 | | | | |

Note: Items that are not meant for human consumption are not mentioned in the information.

Besides $Q_{t,r,p}$, revenue and welfare also depend on the regional product price $z_{r,p}$ (USD $t^{-1}$). The FAO [15] statistics from 2015 were used to project these prices. The author averages the prices of several nations in each RoW region as the RoW producer prices because there are no regional aggregates available. In the absence of producer pricing information for a certain good or location, the author uses the continental (EU-28) or global average. Prices for processed goods like oils are not included in FAOs. The author utilises the FAO [15] trade dataset, which contains information on the values and quantities of goods imported and exported on a global scale, to estimate them. The author determines the value per traded amount and average import and export data to get the selling price of these processed goods. Price predictions using trade values can range from 8% to 300% of the relevant producer price. 80% of the estimates undervalue the price when compared to their actual producer pricing. We reduce the trade-based price estimates by a factor of 1.4 because around half of the data understate the price by 40% or less. Skimmed milk and feed are not included in this estimation method. The feed is only intended for use by cattle, hence it has no monetary value. Skimmed milk, a leftover from the production of butter, is offered for the same price as raw milk. Additionally, although cost estimates call for price estimates for sugarcane and oil palm fruits, their consumer prices are placed at zero because they are not meant for direct human consumption.

### 3.2 Production

Any good that is consumed in Equation (1) must first be created by processes, livestock, or farmland. Additionally, whatever product that is generated is also consumed, whether directly or indirectly, in an optimum model. Due to trade and storage possibilities, this does not, however, mean that a product must be consumed at the same time or location that it was made. These elements are reflected in this model's product balance equation:

$$Q_{t,r,p} = \sum_{c,m}(y^c_{r,c,p} \cdot y^m_{c,m} \cdot C_{t,r,c,m}) + \sum_l(y^l_{r,l,p} \cdot L_{t,r,l}) + \sum_q(y^q_{r,q,p} \cdot P_{t,r,q}) + \sum_{r'}(T_{t,r',r,p} - T_{t,r,r',p}) + S^-_{t,r,p} - S^+_{t,r,p} \forall t,r,p \qquad (6)$$

where $y^c_{r,c,p}$ is the yield (t ha$^{-1}$) of crop $c$'s primary product $p$ from the area $C_{t,r,c,m}$ in region $r$ and period $t$, amplified or reduced by the yield factor $y^m_{c,m}$, depending on the user management system $m$. Livestock yield (t head$^{-1}$) $y^l_{r,l,p}$ can be both positive (primary livestock products) and negative (feed products consumed by livestock). Process yields $y^q_{r,q,p}$ (t t$^{-1}$ produced/processed) are negative for input products and positive for outputs of process $q$. The trade term is positive if region $r$ is a net importer and negative if it is a net exporter in period $t$. Lastly, the author adds products taken from storage $S^-$ (1000 t) to the balance while subtracting products put into storage $S^+$ (1000 t).

### 3.2.1 Crop Production

By restricting the amount of product that is available in Equation (6) and by increasing production costs in Equation (1), crop production has an impact on total welfare. The amount of land that can be used for crops is restricted.

#### 3.2.1.1 Crop Yield

We estimate crop yields $y^c_{r,c,p}$ using FAO [15] yield statistics averaged over the 2010-2014 period per crop $c$ and region $r$. This model takes the baseline yield to be time-independent. If there are no yield data in the dataset being used, the author estimates the yield to be 0. The author makes an exception for the EU-28 because pricing information is accessible at the national level. In this case, the author applies the typical return for this region in the EU. This only applies to apples & beans in Austria, as well as beans in the UK and Northern Ireland.

$$y^m_{c,m} = g^y_m \cdot i^y_m \qquad (7)$$

As illustrated in Equation (7), the management yield factor $y^m_{c,m}$ is composed of two components: the general component $g^y_m$, which includes fertiliser inputs and pest control, and the water component $i^y_m$. $g^y_m$ consists of three basic production systems: conventional, organic, and vast farming. These three systems are either restricted to rainfed production or enlarged by the addition of artificial irrigation. But by definition, large-scale farming can only be rainfed [16]. The model, therefore, includes five different crop management techniques. The typical rainfed management system is regarded as the starting point, with both $g^y_m$ and $i^y_m = 1$, resulting in $y^m_{c,m} = 1$.

Many synthetic inputs are avoided in organic farming processes, which generally provide lower yields. Although crop types, regions, and other factors have a significant impact on the yields from different organically managed farms, we choose a single factor of 0.8 that is independent of these factors for the sake of simplicity [17]. According to Reidsma et al. [16], traditional farms and permaculture systems are just two examples of extensive traditional or organic agricultural methods. Such methods are even further removed from conventional agricultural systems that prioritise output than the organic management methods mentioned above. For instance, inputs for fertiliser and pest control are substantially lower than in organic systems. The author estimates the yield factor of extensive farming based on input cost ranges taken from [16] system definitions, where an intensive system (organic or conventional) costs between 80 and 250 USD per hectare and an extensive system costs less than 80 USD per hectare. These inputs are expensive and only used to increase yield, as there is a lack of yield data in large-scale extensive farms. Assuming a homogeneous cost distribution, with extensive input costs averaging 40 USD and organic intensive costs averaging 165 USD, the author estimate extensive organic system yields to be $\frac{40}{165}$ = 24 % of organic base yields. The author calculates extended yields to be 19% since the author believes organic yields to be 80% of baseline conventional yields. To calculate $i_m^y$ the author average yield changes in artificially irrigated systems vs rainfed systems as:

$$lei_m^y = \frac{\sum_c \frac{y_c^{Irr}}{y_c^{RF}}}{\sum_c 1} \tag{8}$$

where $y_c^{Irr}$ and $y_c^{RF}$ are the irrigated and rainfed yields for nine crops, respectively, as provided by [18]. The unweighted average of these crop-specific yield variables yields $i_m^y$ = 1.38, as indicated in Table 3. Table 4 summarises the derived management-dependent yield factors $y_m^m$. If the global average for a crop's blue water footprint equals zero, as determined since this crop never benefited from irrigation, the author set the irrigated production to zero. This solely applies to the oil palm data.

Table 3: Yield statistics for rainfed and irrigated crops of 1996-2005

|  | Yield(t ha$^{-1}$) | | |
| --- | --- | --- | --- |
| **Crop** | **Rainfed** | **Irrigated** | **Irrigated - Rainfed Factor** |
| Maize | 4.02 | 6.03 | 1.48 |
| Wheat | 2.44 | 3.30 | 1.33 |
| Soybean | 2.21 | 2.46 | 1.12 |
| Apples | 8.99 | 15.90 | 1.78 |
| Rice | 2.56 | 4.66 | 1.74 |
| Rapeseed | 1.65 | 1.22 | 0.75 |
| Coffee | 0.66 | 0.99 | 1.44 |
| Cotton | 1.33 | 2.17 | 1.6 |
| Sugarcane | 58.6 | 71.18 | 1.21 |
| **Average** |  |  | **1.38** |

Table 4: A summary of estimated and determined management factors for conventional (C) and organic (O) systems, rainfed (RF) or artificially irrigated (Irr), extended (E), natural (N), and urban (U) settings

| **Factor** | **C-RF** | **C-Irr** | **O-RF** | **O-Irr** | **E** | **N** | **U** |
| --- | --- | --- | --- | --- | --- | --- | --- |
| Cost $c_m^m$ | 1 | 1.2 | 0.9 | 1.08 | 0.5 | - | - |
| Yield $y_m^m$ | 1 | 1.33 | 0.7 | 1.106 | 0.19 | - | - |
| Water (grey) $w_{wm}^{m,g}$ | 1 | 1.1 | 0.6 | 0.7 | 0 | - | - |
| Water (blue) $w_m^{m,b}$ | 0 | 1.37 | 0 | 1.105 | 0 | - | - |
| Phosphorus $i_r^{m,P}$ | 1 | 1.1 | 0.7 | 0.9 | 0.1 | - | - |
| Nitrogen $i_r^{m,N}$ | 1 | 1.1 | 0 | 0 | 0 | - | - |
| Emissions (crop residues) $e_{c,m,e}^m$ | 1 | 1.38 | 0.8 | 1.104 | 0.19 | - | - |
| Emissions (manure on soil) $e_{c,m,e}^m$ | 1 | 1.1 | 1.2 | 1.2 | 0.1 | - | - |
| Emissions (synthetic N) $e_{c,m,e}^m$ | 1 | 1.1 | 0 | 0 | 0 | - | - |
| Emissions (rice cultivation) $e_{c,m,e}^m$ | 1 | 1.38 | 0.8 | 1.104 | 0.19 | - | - |
| Biodiversity loss $b_m^C$ | 0.9 | 0.95 | 0.8 | 0.85 | 0.65 | - | - |
| Biodiversity loss $b_m^P$ | 0.8 | - | - | - | 0.6 | 0 | - |
| Biodiversity loss $b_m^{F/O}$ | - | - | - | - | - | 0 | - |
| Biodiversity loss $b_m^U$ | - | - | - | - | - | - | 1 |
| TLU ha$^{-1}$ pasture $u_m^m$ | 2 | - | - | - | 1 | 0.3 | - |

Note: Except for biodiversity loss, where the land type is stated (cropland (C), pasture (P), forest (F), other natural lands (O), and urban (U)), and TLU pasture density, all parameters are for cropland.

### 3.2.1.2 Crop Cost

While products are charged by the tonne, crop production costs are measured by crop area $c_{t,r,c,m}$. The cost per area is further affected by a regional baseline factor $C_{r,c}^c$ (USD ha$^{-1}$) and a management factor $C_m^m$. Assuming a balance between production costs and consumer pricing, the author estimates the baseline cost as follows:

$$C_{r,c}^c = z_{r,p} \cdot y_{r,c,p}^c, \tag{9}$$

where $z_{r,p}$ is the consumer price in region $r$ for product p, and $y^c_{r,c,p}$ is the associated product yield from crop $c$. As illustrated in Equation (1), depending on the crop management strategy used $m$, the author adds a management component to this baseline $C^m_m$, comprising a general term and an irrigation term, as given in Equation (10):

$$C^m_m = g^c_m \cdot i^c_m, \tag{10}$$

where $g^c_m$ is the broad word referring to the basic management system and, in particular, the cost of inputs like fertiliser and pesticides, and $i^c_m$ is determined by the irrigation system chosen or not chosen. In terms of yield estimates, the baseline management system with $g^c_m = 1$, $i^c_m = 1$, $C^m_{c,m} = 1$ is a typical rainfed system. The author estimate $g^c_m$ to be 0.9 and 0.6 for organic and extensive systems, respectively. These are rough estimates based on the idea that while these systems use fewer, less expensive inputs, they must offset their lower efficiency with greater labour costs. Compared to an efficient traditional system, costs per hectare may be significantly higher due to unpredictable inputs and labour requirements. Contrarily, organic products usually have higher profit margins, which is not directly reflected in this model's pricing computation. The author maintains the cost of organic systems is lower than the cost of conventional systems to make up for this. In terms of $i^c_m$, irrigation is more expensive than rainfed systems. A cost factor of 1.2 is taken into account for irrigated systems versus the baseline rainfed system based on [19] estimates for maize irrigation in Tanzania compared to the baseline cost estimate for maize in Africa and the Middle East area in this study. The calculated management-dependent cost variables are reported in Table 4.

### 3.2.1.3 Crop Mix

The author ensures more practical production and consumption behaviour through crop-mix restriction because a PEASM without a demand function, as developed for this study, would probably specialise in a few products with a relatively high-profit margin while ignoring many of those, in fact, important products:

$$\sum_m C_{t,r,c,m} = \sum_{t'}(m^c_{t',r,c} \cdot M^c_{t,r,t'}) \quad \forall t, r, c \tag{11}$$

where $m^c_{t',r,c}$ is historical [15] crop-mix data comprising area amounts for crop $c$ in region $r$ and the historical period $1995 \leq t' \leq 2014$, and crop-mix variable $M^c_{t,r,t'}$, which applies these historical amounts in full or in parts to the modelled periods $t$. If any of the historical crops $c$ in region $r$ is to be planted in period $t$ again, $M^c_{t,r,t'} > 0$ is required for at least one $t'$. This allows crop $c$ to be grown but enforces at the same time all other crops $c$ from period $t'$ to be grown according to their historical shares. The function is not enforced for crops without history in region $r$, allowing them to be freely added next to potential historical crop mixes if other constraints do not apply. It is quite uncommon that a crop without historical precedent is planted in a new place in this model, though, as yield data is tied to past values as well and typically only exists based on them.

### 3.2.2 Livestock Production

Livestock production has an impact on welfare both directly and indirectly through the cost of every individual animal as stated in Equation (1) and indirectly through the items, it feeds into the product balance Equation (6) as well as those it receives from it as feed.

### 3.2.2.1 Livestock Yield

When assessing livestock output, cost, and consequences, the author does not make a distinction between various specialised subspecies of livestock, such as broiler and layer hens. For the investigated livestock products, the author uses FAO [15] livestock yield (t head$^{-1}$) statistics for 2014 and, as necessary, converts them to a single animal type that provides both milk and beef or eggs and poultry. As an unweighted average of yield data from FAO sub-regions, the author calculates yield estimates for the RoW areas. In addition to producing basic livestock products, the animal consumes feed. As shown in Table 5, feed consumption is stated as a negative yield term for feed $y^{l,feed}_{r,l}$ (t head$^{-1}$), an aggregate product consisting of many agricultural and animal main products often utilized for feeding. Feed is deducted from the product balance Equation (6) per animal $l$ retained in period $t$ and region $r$ due to its negative sign. To calculate the quantity of non-pasture feed needed per animal in 2014, the author used data from FAO [15]. Since feed is the source of the energy in animal products, the author thinks that higher product tonnage also means higher feed consumption. Based on this supposition, the author determines the following feed requirements for each animal species:

$$y^{l,feed}_{r,l} = -\left(\frac{\sum_{p'} f_{r,p'}}{\sum_p x_{r,p}} \cdot \sum_p y^l_{r,l,p}\right) \tag{12}$$

where $f_{r,p'}$ is the amount of $p'$ documented as feed and $x_{r,p}$ is the number of livestock primary product $p$ produced in region $r$ according to the FAO [15] dataset. As the fraction provides t t$^{-1}$ feed/livestock product, summing over the different product yields $y^l_{r,l,p}$ associated with livestock type $l$ provides the regional $y^{l,feed}_{r,l}$.

Table 5: Most import global feed shares and cost estimation for $c^f$

| Crop | Share [%] | Scaled share [%] | Cost average [USD/t] | Weighted Cost [USD/t] |
|---|---|---|---|---|
| Wheat | 11 | 11 | 304.2 | 33.2 |
| Maize | 51 | 52 | 427.4 | 221.7 |
| Milk | 7 | 7 | 482.4 | 33.0 |
| Cassava | 8 | 8 | 486.2 | 40.9 |
| Rice | 4 | 4 | 1531.7 | 23.6 |
| Barley | 6 | 6 | 297.9 | 19.2 |
| Potatoes | 3 | 3 | 402.4 | 11.3 |
| Sugarcane | 3 | 4 | 135.6 | 4.8 |
| Soybeans | 2 | 2 | 621.9 | 11.1 |
| Sorghum | 2 | 2 | 297.8 | 6.8 |
| Chickpeas | 1 | 1 | 962.2 | 6.3 |
| **Average** | | | | **411.9** |

#### 3.2.2.2 Livestock Cost

The author estimates the cost per animal $l$ and region $r$, $c^l_{r,l}$, assuming a balance between production costs and consumer prices, adjusted for profit loss resulting from goods being used as feed instead of for consumption, as:

$$c_{r,l} = \sum_p (z_{r,p} \cdot y^l_{r,l,p}) - y^{l,feed}_{r,l} \cdot c^f \qquad (13)$$

where $c^f$ is a projection of the global cost per tonne of feed. To do this, the author uses data from the 2014 FAO [15] statistics to determine the global and regional proportions of various feed components and to judge the importance of these proportions. The author excludes less significant food products and scales the selected shares to a total of 100%, estimating the worldwide feed cost $c^f$ = 411.9 USD t$^{-1}$ based on the global average cost per product tonne estimated from FAO [15] data for 2015. This $c^f$ results in very low or negative $c^l_{r,l}$ for cattle in some places, essentially subsidising cattle production in these areas. Cattle raising prices for these areas are manually fixed to a still affordable 300 USD per head because this effect was not intended. Furthermore, the author does not force this feed mix on the model, allowing it to find a more effective feed mix, even though the author uses it to determine average feed or regional livestock production costs.

#### 3.2.2.3 Livestock Mix

As with crop mix, the author employs a livestock mix function to prevent the model from being overly focused on a particular type of livestock, as in:

$$L_{t,r,l} = \sum_{t'} (m^l_{t',r,l} \cdot M^l_{t,r,t'}) \quad \forall t, r, l \qquad (14)$$

where $m^l_{t',r,l}$ is historical FAO [15] livestock-mix data comprising stock numbers (heads) for livestock types $l$ in region $r$ and the historical period $1995 \leq t' \leq 2014$. Variable livestock mix $M^c_{t,r,t'}$ weights these historical data and applies them to the modelled periods $t$, as discussed in subsection 3.2.1.3 for crop mix.

### 3.2.3 Land Constraints

The land that is accessible is what restricts agricultural output, as:

$$a_r = \sum_x LU_{t,r,x} \quad \forall t, r \qquad (15)$$

where $LU^c_{t,r} = \sum_{c,m} C_{t,r,c,m,}, \quad LU^p_{t,r} = \sum_m LUM^p_{t,r,m} \quad \forall t, r \qquad (16)$

and where $a_r$ is the total land area in region $r$ based on FAO [15] statistics from 2010. $LU_{t,r,x}$ is the land use (1000 ha) in the region and period $t$ per land type $x$ (cropland $c$, pasture $p$, forest $f$, other natural land $o$, and urban and other artificial land $u$), and $m$ are the possible management systems in the case of cropland and pasture. $LUM_{t,r,x,m}$ denotes the area (1000 ha) of land type $x$ under management $m$ during period $t$ and region $r$. Similar, the land-use conflict in Equation (15) has two effects on livestock output because livestock needs both a minimum amount of grassland for cattle and hens as well as feed produced from crops, as:

$$\sum_m (LUM_{t,r,m}^p \cdot u_m^m) \geq \sum_{l'} (L_{t,r,l'} \cdot u_{r,l'}^l) \quad \forall t, r \tag{17}$$

where $u_m^m$ is the pasture livestock density (TLU ha$^{-1}$) determined by [16] for different pasture management systems $m$ and described in Table 4, and $u_{r,l'}^l$ is the Tropical Livestock Unit (TLU) for pasture-dependent livestock type $l'$ in region $r$. $u_{r,l}^l$ estimations are based on [20] computations, in which, depending on the model location, we average numerous source values to better reflect diverse realities, as indicated in Table 6.

Table 6: TLU estimations for the region

| Region | Source regions | Cattle | Pigs | Chickens |
|---|---|---|---|---|
| ASI | East and South Asia[Sic], South Asia | 0.575 | 0.225 | 0.01 |
| EU-28 | OECD | 0.9 | 0.25 | 0.01 |
| MAF | Africa South of Sahara, Near East, Near East North Africa, South Africa | 0.613 | 0.213 | 0.01 |
| LAM | South America, Central America, Caribbean [Sic] | 0.667 | 0.233 | 0.01 |
| OCE | OECD | 0.9 | 0.25 | 0.01 |
| NAM | North America | 1 | 0.25 | 0.01 |

The author imposes three restrictions on land use: In the first step of the model, 2015, the author initialises land cover using land shares from the FAO's Climate Change Initiative Land Cover (CCILC) data for the reference year 2010. With a land-use change corridor, the author restricts the maximum growth and loss of any land type $LU_{t,r,x}$ each time step while imposing a minimum proportion of land type by 2050, and the author imposes population-based expansion of urban and other artificial regions $LU_{t,r}^u$. The author utilises Equations (18) and (19) to establish 2015 land-use values as well as to characterize the allowable LUC corridor for the following periods as:

$$lu_{2010,r,x'} \cdot 0.95 \leq LU_{2015,r,x'} \leq lu_{2010,r,x'} \cdot 1.1 \quad \forall r \tag{18}$$

And $LU_{t-1,r,x'} \cdot 0.95 \leq LU_{t,r,x'} \leq LU_{t-1,r,x'} \cdot 1.1 \quad \forall t, r \tag{19}$

where $x' = [c, p, f, o]$ is the set of possible land types $x$, excluding urban, and $lu_{2010,r,x}$ is the reference FAO [15] land use data for land type $x$ in region $r$ in 2010. We arbitrarily choose 5% maximum loss and 10% maximum gain per land type, region, and time step compared to the time step before ($t' - 1$), as these values allow for large-scale changes over time and seem feasible per time step. Additionally, picking a corridor that is too narrow could make the model less adaptable. According to Equations (20) and (21), the author assumes that urban area increases linearly with population numbers in general and that it stays at least at their current level even in the case of a declining population. Furthermore, maximum urban expansion is constrained by the assumption that urban population density will not fall below a certain threshold i.e. 50% of its 2010 value, $\frac{p_{2010,r}}{LU_{2010,r}^u}$:

$$\frac{lu_{2010,r}^u}{p_{2010,r}} \cdot p_{t,r} \leq LU_{t,r}^u \leq \frac{lu_{2010,r}^u}{p_{2010,r}} \cdot p_{t,r} \cdot 2 \quad \forall t, r \tag{20}$$

$$LU_{2010,r}^u \leq LU_{t-1,r}^u \leq LU_{t,r}^u \quad \forall t, r \tag{21}$$

where $p_{t',r}$ is the total population (millions) in region $r$ and period $t' = [2010, t]$, as projected in the SSPDB for SSP 2 [21].

### 3.2.4 Processes

In this paradigm, the third way to manufacture products is to convert main agricultural and livestock products into secondary ones. This affects product balance Equation (6) twofold, by adding process outputs and removing process inputs through yield $y_{r,q,p}^q$, and welfare Equation (1) due to process costs $c_{r,q}^q$.

### 3.2.4.1 Process Yield

Process variable $P_{t,r,q}$ counts process processing 1000t of the respective input product. Hence, process requirements for all input products $i$ is $y^q_{r,q,i} = -1$. The model has 17 processes in total, 11 of which convert the most often used feeding items, as stated in Table 5, into feed at 100% efficiency. Table 7 shows how the other six processes provide geographically varied yields of seven consumer commodities as well as feed as a byproduct.

Table 7: Main product processes

| Input | Output(s) | Input | Output(s) |
|---|---|---|---|
| Oil palm fruits | Palm oil | Rapeseeds | Rapeseed oil, feed(cake) |
| Milk | Butter, Skimmed Milk | Olives | Olive oil |
| Soybeans | Soy oil, feed(cake) | Sunflower seeds | Sunflower oil, feed(cake) |

For the oil processes, the author determines $y^q_{r,q,p}$ as shown in Equation (22):

$$y^q_{r,q,p} = \frac{\hat{p}_{2013,r,p}}{\hat{p}_{2013,r,i}} \tag{22}$$

where $\hat{p}_{2013,r,p}$ is the production quantity of the process output product and $\hat{p}_{2013,r,i}$ is the processed quantity of the process input product, based on [15] data for 2013 in region $r$. The commodity balances dataset contains no information on the input crop, oil palm fruit, for palm oil. To address this void, the author uses the [15] crops dataset and substitutes the production quantity per area. To estimate the processing share of crop production $p'_{t,r,c}$, the author compares available crop production and processing data for other oil crops. Based on this, the author estimate that, on average, 82% of produced oil palm fruits are processed and substitute $\hat{p}_{2013,r,oil\ palm\ fruit} = 0.82 \cdot p'_{t,r,oil\ palm\ fruit}$ in Equation (22) to estimate $y^q_{r,q,p}$ for palm oil production.

The author presumes uniform global efficiency levels and estimates it for the butter process using the procedures provided by [22] and summarised in Equations (23) and (24):

$$y^b = \frac{(1-y^{m'}) \cdot f^c}{f^b} \tag{23}$$

With $y^{m'} = \frac{f^c - f^m}{f^c - f^{m'}}$ (24)

where $f^x$, with $x = [m, m', c, b]$, is the fat content of unprocessed milk $m$, skimmed milk $m'$, cream $c$, as intermediate production step, and butter $b$. Assuming $f^m = 4\%$, $f^{m'} = 1.5\%$, $f^c = 40\%$, and $f^b = 80\%$, we estimate $y^{m'} = 0.935$ t t$^{-1}$ skimmed milk and $yb = 0.0325$ t t$^{-1}$ butter yield per processed ton of milk.

### 3.2.4.2 Process Cost

Like livestock and crop production costs, the author estimate process costs using the expected consumer price $z_{r,p}$ per tonne of process output product $\dot{p}$, as discussed in section 3.1, the process yield efficiency $y^q_{r,q,p}$ (t t$^{-1}$), and the cost of the processed input $c^i_{r,i}$. As the process "yield," or input requirement, for the input product is defined as $y^q_{r,q,p} = -1$, the process cost is represented as:

$$c^q_{r,q} = \sum_p (z_{r,p} \cdot y^q_{r,q,p}), \text{ with } p=[\dot{p},i] \tag{25}$$

As a result, consideration is given to both the input cost and the price of the final product. Additionally, both product pricing and yield efficiency are taken into account in the distinctive case of the butter process, which creates two consumer commodities, namely butter and skimmed milk. As opposed to that, food is free. As a result, it raises the cost of the operation rather than serving as a byproduct of the only output product. In rare circumstances, cost correction causes process costs to be negative for certain locations and processes $q'$. For these, the author estimate process cost as a worldwide cost average across areas with $c^q_{r,q'} > 0$.

### 3.2.5 Trade and Storage

A good produced in one area can be used or processed in another through trade, whereas a good stored for a long can do the same.

### 3.2.5.1 Trade

The product balance Equation (6) is the only trade limitation, with the upper limit for exports from region $r$ being the entire available product quantity in $r$ and the upper limit for imports to $r$ being the sum of available goods in all other regions $r'$. For the sake of simplicity, the author set commerce between two locations to 0 and does not use historical trade routes as a baseline. The author calculates trade costs as a percentage of import value (USD t$^{-1}$), according to the UN Conference on Trade and Development. For the sake of simplicity, the author assumes a global average of 15% of the regional product price $z_{r,p}$ for all commerce, regardless of trading partners. There is no $z_{r,p}$ for items that are not intended for direct human consumption. The author is assuming a flat transportation cost for these.

### 3.2.5.1 Storage

In this model, storage capacity is limitless, therefore $S^L_{t,r,p}$ is only restricted by product availability and minimal consumption, which limit the quantity $S^+_{t,r,p}$ that may be added to storage in region $r$ and time $t$, as:

$$S^L_{t,r,p} = S^L_{t-1,r,p} + S^+_{t,r,p} - S^-_{t,r,p} \quad \forall t, r, p \tag{26}$$

where $S^L_{t-1,r,p}$ is the storage level in the previous period (0, if $t$ = 2015). Furthermore, to reflect product perishableness, products can only be stored for one period, as:

$$S^-_{t,r,p} \geq S^+_{t-1,r,p} \quad \forall t, r, p \tag{27}$$

Storage combines two different cost types. For $S^+_{t,r,p}$ and $S^-_{t,r,p}$, the author considers transport to and from storage. For $S^L_{t,r,p}$, the author assumes rent and cooling to be important cost factors. In each of these examples, the author estimate that these costs are merely a fraction of the costs of international commerce, which combine these issues on a bigger scale, such as greater distances. As a result, the author estimates the cost of the storage transit procedure to be 1% of $S^L_{t,r,p}$ and 3% of the storage term.

## 4. Result

Equation (1)'s final term internalises externality costs via taxes. GHG emissions, BDL, deforestation, usage of blue and grey waters, artificial nitrogen fixation, and phosphorus pollution are some of these externalities. Each externality's worldwide cumulative level from 2015 to 2050, in addition to its cost, serves as a sustainability effect indicator for the agricultural system model. The author defines the baseline run by setting all taxes to zero by default. To highly encourage system choices that minimise the taxed externality, the author tax one of the externalities with an unreasonably high cost per externality unit in six additional runs. This allows us to study the lowest levels each indicator may go to and how this affects other indicators. The author run the model seven times. The first run, scenario 0, with $t_e = 0 \, \forall_e$, defines the base to which the author compares all other scenario runs. In scenario 1, the author set $t_{CO2eq}$ = 10000 USD t$^{-1}$ CO$_{2eq}$, in scenario 2–6, the author set $t_e$ = 1000000 USD/externality unit (second run: biodiversity loss (1000 ha$^{-1}$ carrying capacity lost), third run: deforestation (1000 ha$^{-1}$), fourth run: freshwater use (1000m$^{-3}$), fifth run: synthetic nitrogen (t$^{-1}$), sixth run: phosphorus (t$^{-1}$)). The resulting $R^E_j$ values are shown in Figures 1, 2, 3, 4, 5, and 6, respectively.

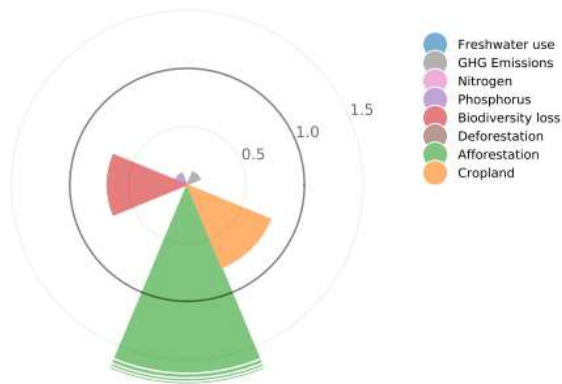

Figure 1: $R^{co2eq}_j$ values from scenario first run

In Scenario 1, where the author imposes a price on GHG emissions, afforestation rises to as much as 2.7 times its baseline level, while all other indicators fall in tandem with GHG emissions, as seen in Figure 1. Most notably, deforestation, freshwater use, and synthetic nitrogen application all drop to zero or very close to zero, while phosphorus drops to 11% of the base level, which appears to be the lowest level in this model possible, as it does not drop lower even in scenario 6, where phosphorus is taxed, as shown in Figure 6. Aside from Scenario 2, where BDL is directly charged, the loss level in Scenario 1 is also at its lowest, lying at 68% when compared to the base scenario. Cropland declines by 23%, the smallest indicator change in this scenario but the biggest cropland reduction predicted in comparison to other scenarios. Scenario 2, depicted in Figure 2, reduces BDL to the lowest amount among the six scenarios, at 26% of the base scenario. Forests see no change since there is no deforestation or afforestation in this scenario. As in Scenario 1, 23% of farmland in the basic scenario is freed up. Aside from Scenario 1, GHG emissions reach their lowest level at 39%. While nitrogen, phosphorus, and freshwater levels are not at their lowest conceivable point when compared to other scenarios, they all drop very low relative to the baseline, at 5, 16, and 18%, respectively.

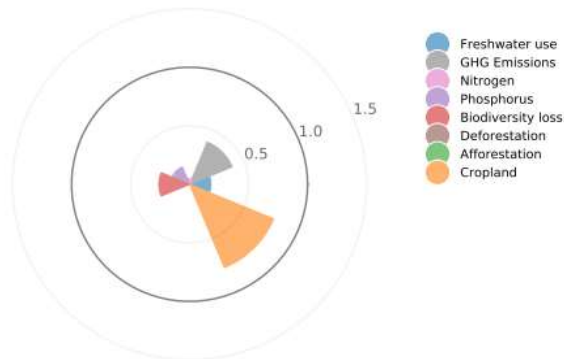

Figure 2: $R_j^{BDL}$ values from scenario second run

Taxing deforestation in Scenario 3 (illustrated in Figure 3) reduces this indicator to zero, whereas most others either decrease or rise. Cropland declines by 7%, but biodiversity loss decreases by 10% and GHG emissions decrease by 13%. Water, nitrogen, and phosphorus footprints, on the other hand, increase beyond baseline levels by 102, 117, and 110%, respectively. And, while deforestation decreases, afforestation increases by 75% relative to the baseline scenario. Only scenario 2 decreases.

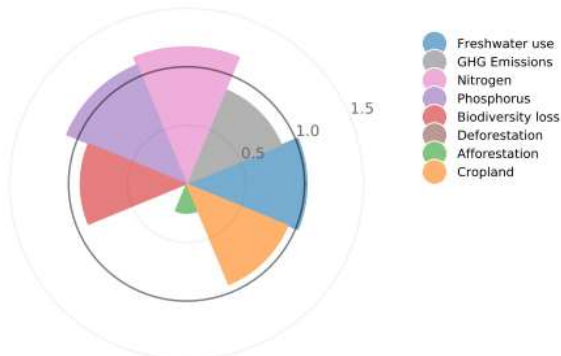

Figure 3: $R_j^{Defor}$ values from scenario third run

Scenario 4 taxes blue and grey water consumption and drives it to its lowest potential level, close to zero as shown in Figure 4. Simultaneously, all other indices decrease, except for afforestation, which reaches its peak, scenario 1 apart, at 179% of the baseline level. In the same period, farmland decreased by 11% and deforestation decreased by 39%. GHG emissions are reduced by nearly half, to 53%. BDL is reduced to 70% of the baseline value. Phosphorus and nitrogen levels fall, even more, reaching 27% and 10% of their respective baselines.

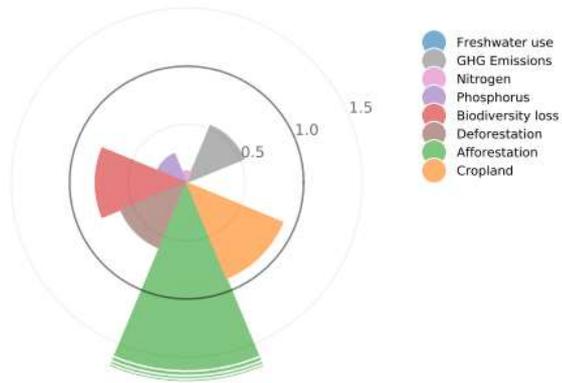

Figure 4: $R_j^{Water}$ values from scenario fourth run

The taxation situation for synthesised nitrogen is depicted in Figure 5. Few indications shift when this fertiliser input falls to zero. GHG emissions are reduced by 14%, one of the smallest reductions among all scenarios. Water use is also decreasing, but only by 16%. Cropland and BDL are almost unchanged at 100% and 95% of baseline levels, respectively. In the same period, phosphorus input rises by 5%, while deforestation rises by 30%. However, this is somewhat offset by a 20% rise in afforestation.

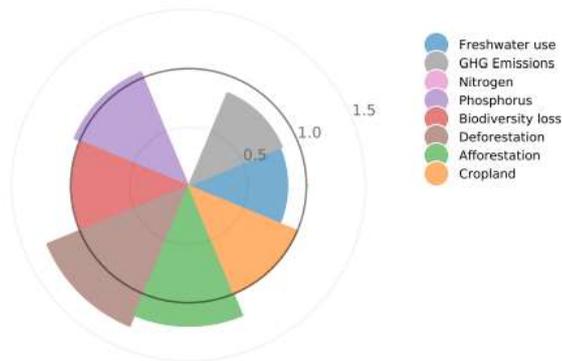

Figure 5: $R_j^N$ values from scenario fifth run

Finally, Figure 6 depicts the phosphorus pollution tax scenario, which reduces phosphorus application to 11% of its base value. Nitrogen and water use are reduced even more to zero. Along with phosphorus, GHG emissions and BDL are reduced by 52% and 70%, respectively. While agriculture is reduced by 23% and afforestation grows by 30%, deforestation increases by 24%, diminishing the favourable climatic impact of the increased forest.

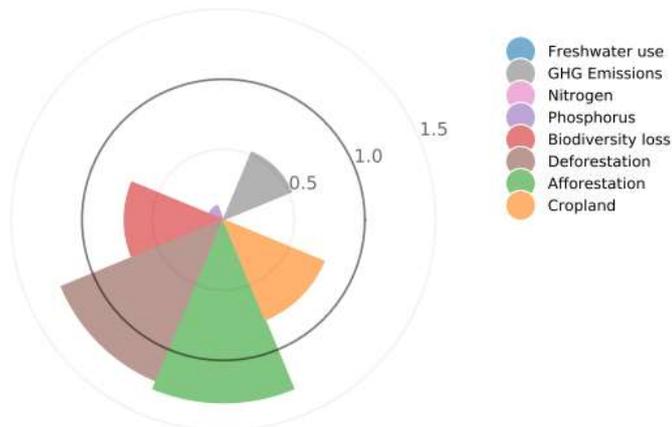

Figure 6: $R_j^P$ values from scenario sixth run

Figure 5 shows that the primary advantage of eliminating synthetic nitrogen has few co-benefits in this scenario. The nitrogen indicator, on the other hand, benefits from the majority of the other taxes. As seen in Figure 7, this is because many tax scenarios shift crop production primarily or entirely too vast farming where synthetic nitrogen application is prohibited, except for deforestation, which does not benefit from the lower area efficiency. In the long term, the nitrogen tax shifts crop output to an otherwise uncommon management technique irrigated organic farming, which delivers the second greatest yields while restricting nitrogen application. Similarly, as seen in Figure 3, deforestation has few evident significant correlations with other variables. Only afforestation exhibits changes comparable to a 100% reduction in deforestation. This makes obvious sense since the model compensates for the lack of accessible land for agricultural or pasture development in the basic scenario by preventing other regions from becoming afforested. These are presumably less efficient than those deforested in the basic scenario, but they are required for profit maximisation. The crop-management distribution in Figure 7 reflects the reduced efficiency of these now-un afforested regions.

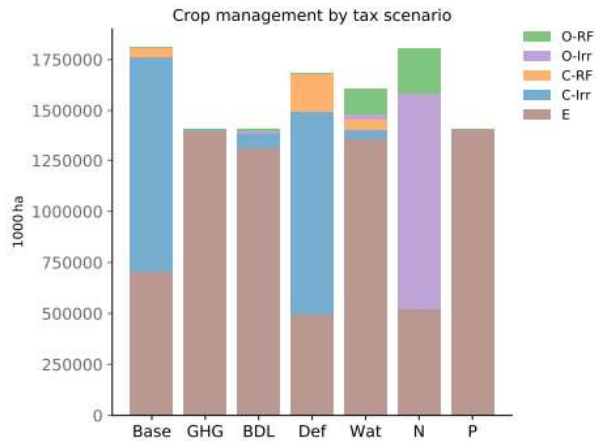

Figure 7: Annual average area (1000 ha) of agriculture managed under the various accessible management systems from 2015 to 2050, per tax scenario

While farmland is reduced overall, traditional farming rises substantially, particularly C-RF, whose absolute acreage climbs by more than 300%, resulting in a 380% increase in its relative proportion of cropland. This might also explain why the linkages between climate protection and biodiversity integrity are so poor. On the one hand, limiting deforestation directly reduces GHG emissions and BDL from land use change and habitat destruction, both of which are penalised more severely in this model than the same level of afforestation would be, making it more efficient to not remove forests rather than replanting what has been removed. Moving to greater efficiency systems, on the other hand, is related to increased emissions over time and worse ecosystem quality in these locations, as indicated in Table 4. While nitrogen may be significantly lowered by switching from one high-yield management method to a somewhat less effective one, phosphorus, the second analysed nutrient, cannot. It requires a larger transformation. One distinction between the two is that this model focuses on synthetic nitrogen pollution from the atmosphere (as opposed to nitrogen pollution from other sources, such as manure), which is completely prohibited by some management systems, whereas phosphorus pollution is an issue that the author assumes affects all systems, albeit to varying degrees. As a result, the phosphorus footprint cannot be lowered to zero, and any reduction has a significant impact on yield efficiency. Because the tax impacts all management systems, it incentivizes decreasing cropland and expanding its wide management share, as seen in Figure 7. This managerial shift also explains the well-articulated synergy. Crop production is the only source of freshwater consumption in this model, which is decreased to zero in heavily managed fields, as does nitrogen application.

Given the low production efficiency of extensive farming in this model, it is not surprising that consumption patterns shift dramatically in this scenario, as seen in Figures 8 and 9. Crop and livestock product consumption is kept to an absolute minimum, while the model focuses on high caloric but low-value oil products, notably rapeseed oil, and milk, which refines and raises the value of feed crops.

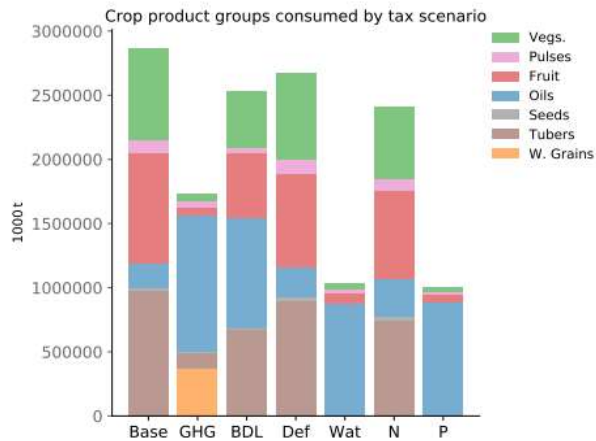

Figure 8: Annual average amount (1000 t) of direct human consumption of agricultural primary and secondary products from 2015 through 2050

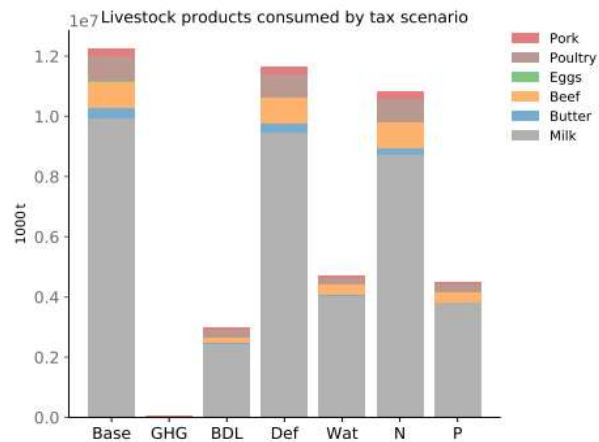

Figure 9: Annual average quantity (1000 t) of direct human consumption of cattle main and secondary products from 2015 through 2050

High deforestation paired with high afforestation may result from a desire to reduce agriculture to reduce the phosphorus effect while also focusing on the most efficient regions. As a result, while regions with high rapeseed yields are deforested, those with lesser yields are afforested to limit farmland damage. This looping diminishes the otherwise good impact of cropland decrease on GHG emissions and BDL. Despite these restrictions, there are beneficial linkages between P, GHG, and BDL reductions in the model as it is now configured. Taxing blue and grey water consumption, as shown in Figure 4, has comparable outcomes to the phosphorus scenario, with a large shift toward intensive crop management, about tripling highly managed farmland and confining crop production almost entirely to this very water-efficient system. Cropland and nitrogen decrease are less evident, but so is deforestation, which is higher than afforestation. The phosphorus indicator now gains from the water-tax-induced management change, just as water did from the phosphorus tax. As a result, consumption moves to the same high-oil, high-milk diet seen in Figures 8 and 9.

The GHG emissions tax has the opposite impact. Here, consumption shifts almost entirely away from cattle goods. This makes it reasonable given that these products have numerous GHG implications, both through feed and direct livestock emissions. As a result, it is not unexpected that when confronted with a high $CO_{2eq}$ price, the model seeks out more efficient paths that do not involve animal consumption. This seeming high leverage is most likely due to a rather abrupt diet change between the basic scenario and scenario 1. Whole grain products are being transferred to human consumption since they are no longer the most effective feed. Cropland can be reduced in this manner despite the poorer yield efficiency of practically solely comprehensive crop management. As long as sufficient land is available without the requirement for emission-intensive land use change, transitioning to extensive management is logical, considering the parameters proposed in this study. While yields fall by 80%, emissions from two sources practically disappear. Further, emissions decrease and increase with production, making vast farming competitive in this area as well. Furthermore, as with all scenarios that change to comprehensive management, the effects of water, phosphorus, and nitrogen are minimised. Furthermore, biodiversity benefits from the conversion of primarily traditionally managed grasslands to woods, as seen in Figures 10 and 11, resulting in an increase in ecosystem quality despite the lower credit. The tax-motivated halting of deforestation is an added benefit for biodiversity conservation. This behaviour will, of course, alter with a more complicated model that takes into account the complex features of different land types and areas, as well as the labour-intensive procedures and crop interactions that are particularly crucial in organic and extensive farming.

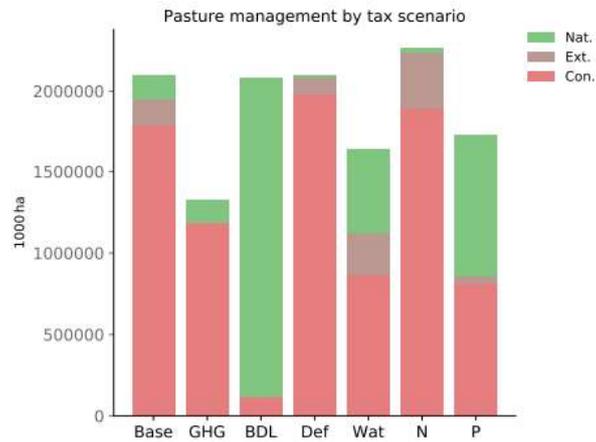

Figure 10: Annual average area (1000 ha) of pasture under extensive (Ext.) or conventional (Con.) management, or without management, from 2015 to 2050

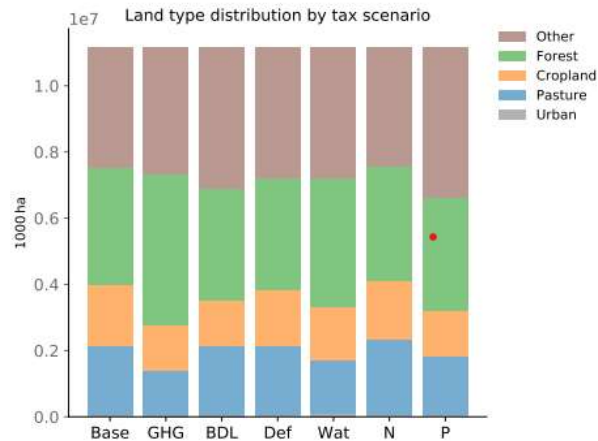

Figure 11: Global agriculture, forest, pasture, urban, and other natural lands annual average area (1000 ha) from 2015 to 2050.

Figure 2 depicts the most dramatic overall changes in the tax on BDL. Aside from cropland, all indicators go below 50%, with five of them falling below 20% of baseline values. Figure 7 shows that, as with runs 1, 4, and 6, this is associated with a significant rise in comprehensive crop management. Furthermore, pasture management is changing dramatically, with the majority of it shifting to natural management, which improves ecosystem quality but decreases cattle carrying capacity. The co-evolution of GHG emissions and BDL may be much more evident if the model was solved in such a manner that the rise in other natural lands, as illustrated in Figure 11, was instead shifted to forests. This should be easy to do in this model because both land types have had the same levels monitored and would carry with them the forest's anticipated increased $CO_2$ sequestration rate. A little tweak in parameter selection might change this and boost the BDL-GHG synergy without harming others. Table 10 lists all synergies and conflicts between the researched planetary impacts, where the percentage represents the proportional amount of the change in each indicator to the change in the taxed externality. As one measure of land-system change, afforestation exhibits both the greatest synergy and the most conflict. While the extra indicators for land-system change are varied in terms of synergies and conflicts, cropland demonstrates consistent synergies in the range of 20-30% with all externality reductions except nitrogen. The interactions between GHG, BDL, water and Phosphorus are most noticeable, with cropland and Nitrogen also benefiting despite the minimal synergies shown by the Nitrogen scenario. These outcomes are, of course, extremely reliant on the model's complexity and parameter selection. For example, in some situations, the consumption combination of predominantly milk and oil reveals a present flaw in the model. While it strives to mirror actual production mixes, consumption can quickly become extremely specialised, which should be handled by including a consumption mix corridor or demand. The seeming adaptability of intensive crop management as a remedy for any type of planetary effect from land usage is also quite visible. While this may be true in theory, the parameter estimates for extensive farming are probably overly optimistic due to a lack of data on large-scale extended operations and a lack of restrictions to represent particular needs such as crop rotation or multiple cropping.

Table 10: Synergies and conflicts between planetary effects indicated as change relative to taxed externality change. When the taxed externality and a sustainability indicator increase or diminish simultaneously, synergies emerge, and percentages represent the relative growth. It is regarded as a synergy for afforestation if it expands while the taxed externality declines, and vice versa.

| Tax | Synergy | Conflict |
|---|---|---|
| BDL | Deforestation (136%), N (129%), P (101%), water (111%), GHG (83%), cropland (31%) | Afforestation (134%) |
| GHG | Afforestation (198%), water (114%), N (113%), deforestation (113%), P (101%), BDL (36%), cropland (26%) | - |
| Water | N (90%), afforestation (79%), P (73%), GHG (47%), deforestation (39%), BDL (22%) cropland (11%) | - |
| Def | GHG (13%), BDL (10%), cropland (7%) | Afforestation (75%), N(17%), P (10%), water (2%) |
| Phosphorus | N (112%), water (112%), GHG (54%), BDL (34%), afforestation (34%), cropland (25%) | Deforestation (26%) |
| Nitrogen | Afforestation (20%), water (16%), GHG (14%), BDL (1%) | Deforestation (30%), P (5%) |

The framework developed in this study is the first step toward a broader model for calculating linkages between planetary impacts, and is constrained by numerous such simplifications, both in data aggregation and model structure: land type classification is quite wide, and the model allows for changes from any to any type within the prescribed land use change corridor without respect for significant real-world constraints such as climate or soil appropriateness. Furthermore, as intended, the RoW areas are relatively vast, with widely distinct climatic zones and economic realities. Sorting them into subregions, while also integrating sections of the world previously disregarded by the model, may improve the model's reliability. Water should be included as a resource in production calculations, possibly severely restricting production in general and specialised management approaches in various regions of the world. Green water might be studied alongside blue and grey water, both as a standalone aspect and as part of a combined total footprint. Forests are not currently seen as producing zones with many management choices and consumable output. Similarly, the production and use of biofuels are now overlooked. Emission estimates, as well as the global feed mix and cost, must be reassessed for animal production.

## 5. Discussion

How relevant are the interactions between various planetary effects of land use, such as climate change, biosphere integrity, nitrogen and phosphorus biogeochemical fluxes, freshwater use, and land system change, as well as their synergies and conflicts? Consumption at a level that, in theory, provides for both food security and zero hunger is a crucial restriction for the model to reflect the tension between the multiple SDGs that have been highlighted. Additionally, the author estimate proxies rather than modelling all of the examined planetary impacts in the same units as [3] have defined. For the climate system, they use either radiative forcing (wm-2) or atmospheric $CO_2$ concentration (ppm). Instead, the author models emissions from land use worldwide (t CO2eq). Rockstrom et al. [3] used the extinction rate (million-1 yr-1) to measure the integrity of the biosphere, whereas the author measured it using regions (1000 ha) with different ecosystem quality levels in comparison to wild systems. The author uses the same approach as [3] to calculate atmospheric Nitrogen fixation for agricultural use using estimations from synthetic fertilisers. Although the author doesn't directly calculate phosphorus run-off into the ocean, the author can utilise anticipated agricultural inputs as a substitute. For blue and grey water, the author employs the same indicator—water consumption ($m^3$)—as [3]. Since 15% of the world's ice-free land serve as the threshold for land-system change, the author continues to utilise the same indicator. The author also examines afforestation and deforestation.

## 6. Conclusion

In this paper, the author aimed to evaluate the connections between planetary consequences and the land-use industry. To achieve SDG-2 (zero hunger), the author created a new agricultural-sector model that calculates GHG emissions, BDL, nitrogen and phosphorus biogeochemical fluxes, blue and grey freshwater consumption, and land-system change. By taxing one of the planetary impacts and comparing its change to the changes in the others and associated sustainability indicators, the author was able to determine how significant the synergies and conflicts between impact reductions are. As shown in Table 10, the author found strong synergies between reductions in GHG emissions, BDL, water use, and phosphorus pollution within the framework of this model. Reducing any one of these will reduce the others by at least 20%, frequently 30% or more, and as much as 100% relative to the change in the taxed externality. Other metrics of land-system change, such as deforestation and afforestation, are less closely related, although eliminating any of these four has a favourable impact on nitrogen application and farmland loss. Between the base scenario and the investigated ones, there is expected to be a significant change in crop management systems that will influence many of these relationships. All of these related systems will specialise to varying degrees in intense farming on a small amount of farmland. The fundamental goal of stopping deforestation, however, has weak synergies with some measurements and minor conflicts with others. Although this effect metric gains more from declines in other indicators, it is also true for nitrogen application. These results can only be considered as an indication of likely relationships between planetary impacts from the land-use sector and potential sustainable measures to mitigate these impacts due to the model's multiple simplifications. Nevertheless, this study advances our understanding of the problem while highlighting areas and issues for further investigation, such as the need for extensive farming.

**Declarations**

**Funding:** There was no funding for this study in any kind.

**Competing interests/Conflict of interest:** The authors declare that they have no competing interests.

**Author's contributions:** All authors contributed equally and significantly in writing this article. All authors read and approved the final manuscript.

**Rights and permissions:** This article is distributed under the terms of the Creative Commons Attribution.

**Availability of data and material:** The data used to support the results of the study may be obtained from the corresponding author.

**Code availability:** Coding is available for this article whenever required.

**Acknowledgment:** The authors are thankful to the learned reviewer for his valuable comments.

**Compliance with ethical standards:** Not applicable.

**References**

[1] L. Hug, M. Alexander, D. You, and L. Alkema, "National, regional, and global levels and trends in neonatal mortality between 1990 and 2017, with scenario-based projections to 2030: a systematic analysis," *The Lancet Global Health*, vol. 7, no. 6, pp. e710–e720, 2019, doi: 10.1016/S2214-109X(19)30163-9.

[2] P. J. Crutzen, "Geology of mankind," *Nature*, vol. 415, no. 6867. p. 23, Jan. 03, 2002, doi: 10.1038/415023a.

[3] J. Rockström et al., "Planetary boundaries: Exploring the safe operating space for humanity," *Ecology and Society*, vol. 14, no. 2, 2009, doi: 10.5751/ES-03180-140232.

[4] IPCC, "Impacts of 1.5°C Global Warming on Natural and Human Systems," in *Global Warming of 1.5°C*, 2022, pp. 175–312.

[5] D. W. O'Neill, A. L. Fanning, W. F. Lamb, and J. K. Steinberger, "A good life for all within planetary boundaries," *Nature Sustainability*, vol. 1, no. 2, pp. 88–95, 2018, doi: 10.1038/s41893-018-0021-4.

[6] J. Randers et al., "Achieving the 17 Sustainable Development Goals within 9 planetary boundaries," *Global Sustainability*, vol. 2, 2019, doi: 10.1017/sus.2019.22.

[7] R. R. Palatnik, F. Eboli, A. Ghermandi, I. Kan, M. Rapaport-Rom, and M. Shechter, "INTEGRATION of GENERAL and PARTIAL EQUILIBRIUM AGRICULTURAL LAND-USE TRANSFORMATION for the ANALYSIS of CLIMATE CHANGE in the MEDITERRANEAN," *Climate Change Economics*, vol. 2, no. 4, pp. 275–299, Nov. 2011, doi: 10.1142/S2010007811000310.

[8] H. Valin, P. Havlík, A. Mosnier, M. Herrero, E. Schmid, and M. Obersteiner, "Agricultural productivity and greenhouse gas emissions: Trade-offs or synergies between mitigation and food security?," *Environmental Research Letters*, vol. 8, no. 3, 2013, doi: 10.1088/1748-9326/8/3/035019.

[9] A. Mosnier, "Tracking indirect effects of climate change mitigation and adaptation strategies in agriculture and land use change with a bottom-up global partial equilibrium model by," *University of Natural Resources and Life*, no. March, 2014, Accessed: Oct. 03, 2022. [Online]. Available: https://epub.boku.ac.at/obvbokhs/content/titleinfo/1931250/full.pdf.

[10] L. Panichelli and E. Gnansounou, "Impact of agricultural-based biofuel production on greenhouse gas emissions from land-use change: Key modelling choices," *Renewable and Sustainable Energy Reviews*, vol. 42. Elsevier Ltd, pp. 344–360, 2015, doi: 10.1016/j.rser.2014.10.026.

[11] P. Havlík *et al.*, "Global land-use implications of first and second generation biofuel targets," *Energy Policy*, vol. 39, no. 10, pp. 5690–5702, 2011, doi: 10.1016/j.enpol.2010.03.030.

[12] K. Riahi *et al.*, "The Shared Socioeconomic Pathways and their energy, land use, and greenhouse gas emissions implications: An overview," *Global Environmental Change*, vol. 42, pp. 153–168, 2017, doi: 10.1016/j.gloenvcha.2016.05.009.

[13] T. Hasegawa, P. Havlík, S. Frank, A. Palazzo, and H. Valin, "Tackling food consumption inequality to fight hunger without pressuring the environment," *Nature Sustainability*, vol. 2, no. 9, pp. 826–833, 2019, doi: 10.1038/s41893-019-0371-6.

[14] "What's In The Foods You Eat Search Tool : USDA ARS." https://www.ars.usda.gov/northeast-area/beltsville-md-bhnrc/beltsville-human-nutrition-research-center/food-surveys-research-group/docs/whats-in-the-foods-you-eat-search-tool/ (accessed Oct. 03, 2022).

[15] "FAOSTAT." https://www.fao.org/faostat/en/#home (accessed Oct. 03, 2022).

[16] P. Reidsma, T. Tekelenburg, M. Van Den Berg, and R. Alkemade, "Impacts of land-use change on biodiversity: An assessment of agricultural biodiversity in the European Union," in *Agriculture, Ecosystems and Environment*, 2006, vol. 114, no. 1, pp. 86–102, doi: 10.1016/j.agee.2005.11.026.

[17] T. De Ponti, B. Rijk, and M. K. Van Ittersum, "The crop yield gap between organic and conventional agriculture," *Agricultural Systems*, vol. 108, pp. 1–9, 2012, doi: 10.1016/j.agsy.2011.12.004.

[18] M. M. Mekonnen and A. Y. Hoekstra, "The green, blue and grey water footprint of crops and derived crop products," *Hydrology and Earth System Sciences*, vol. 15, no. 5, pp. 1577–1600, 2011, doi: 10.5194/hess-15-1577-2011.

[19] I. Carruthers, "Economics of Irrigation," in *Sustainability of Irrigated Agriculture*, 1996, pp. 35–46.

[20] P. Chilonda and J. Otte, "Indicators to monitor trends in livestock production at national, regional and international levels," *Livestock Research for Rural Development*, vol. 18, no. 8, p. 2006, 2006, Accessed: Oct. 03, 2022. [Online]. Available: https://www.researchgate.net/profile/Joachim-Otte-2/publication/286826159_Indicators_to_monitor_trends_in_livestock_production_at_national_regional_and_international_levels/links/626f8ee263e2e65684ba50cb/Indicators-to-monitor-trends-in-livestock-productio.

[21] S. KC and W. Lutz, "The human core of the shared socioeconomic pathways: Population scenarios by age, sex and level of education for all countries to 2100," *Global Environmental Change*, vol. 42, pp. 181–192, 2017, doi: 10.1016/j.gloenvcha.2014.06.004.

[22] J. C. Lambert, "[Village milk processing]. [French]," *Etude FAO: Production et Sante Animales (FAO). no. 69.*, p. 73, 1988, doi: 10.3/JQUERY-UI.JS.